\documentclass[aip,rsi,reprint]{revtex4-1} 

\usepackage{graphicx}
\graphicspath{{figures/}}
\usepackage{siunitx}
\usepackage{epstopdf}
\DeclareSIUnit{\dBm}{dBm}
\usepackage{xspace}
\usepackage{amsmath}
\newcommand{\RFM}{\SI{2856}{\MHz}\xspace}
\newcommand{\RFG}{\SI{2.856}{\GHz}\xspace}

\usepackage{xcolor}

\draft 

\begin{document}


\title{Temporal synchronization of GHz repetition rate electron and laser pulses for the optimization of a compact inverse-Compton scattering x-ray source} 



\author{M.R. Hadmack}
\author{E.B. Szarmes}
\author{J.M.J. Madey}
\author{J.M.D. Kowalczyk}

\affiliation{Department of Physics and Astronomy, University of Hawai`i at M\=anoa, Honolulu, Hawai`i}



\date{\today}

\begin{abstract}
The operation of an inverse-Compton scattering source of x-rays or gamma-rays requires the precision alignment and synchronization of highly focused electron bunches and laser pulses at the collision point.  
The arrival times of electron and laser pulses must be synchronized with picosecond precision.  
We have developed an RF synchronization technique that reduces the initial timing uncertainty from \SI{350}{\ps} to less than \SI{2}{\ps}, greatly reducing the parameter space to be optimized while commissioning the x-ray source.
We describe the technique and present measurements of its performance.
\end{abstract}


\maketitle 

\section{Introduction}
A compact, high-brightness x-ray source based on Compton-backscattering is presently under development at the University of Hawai`i\cite{hadmack2012phd,madey2013oce}.
Picosecond \SI{3000}{\nm} laser pulses at \RFG repetition rate from the Mark~V Free-electron Laser are collided head-on with the \SI{40}{\MeV} electron bunches that drive the laser, resulting in a beam of \SI{10}{\keV} x-rays.  
Since the same electron beam is used to drive both the x-ray scattering interaction and the FEL we can take advantage of the fixed phase relationship between the \si{\GHz} rep-rate electron micro-bunches and optical pulses.  
Due to the Compton cross section of the electrons a very tight focus is required at the interaction point to maximize luminosity\cite{madey2008high}.
While co-alignment of electron and laser beams focused to \SI{60}{\um} diameter spot sizes is a central challenge to the project, it is equally important that the arrival times of the micro-pulses at the interaction point are synchronized.  
Efficient x-ray scattering requires that the interaction be confined to the \SI{2}{\mm} long confocal region of the laser at the designated interaction point.  
Since laser micro-pulses are less than \SI{3}{\ps} in duration, picosecond level synchronization is necessary.  
Given the \SI{350}{\ps} inter-pulse spacing corresponding to the \RFG repetition rate, this synchronization requirement corresponds to an RF phase synchronization of less than \ang{1}.

Many factors confound efforts to measure the arrival time of micro-pulses at the interaction point (IP):  high radiation backgrounds, the broad bandwidth of optical transition radiation, low optical transition radiation (OTR) intensity at \SI{3000}{\nm}, the detection of low level IR light, detector bandwidth, etc.  
The approach described in this paper compares the RF phase of the visible OTR from a copper mirror inserted into the electron beam with the visible coherent harmonic radiation (CHR) generated in the FEL at harmonics of the laser wavelength\cite{bamford1990aa}.
Like incoherent spontaneous radiation from individual electrons, the CHR is caused by deviations from a pure sinusoidal trajectory in a plane-polarized undulator. 
The coherence of the CHR is the result of bunching on the scale of the optical wavelength by the fundamental FEL amplification process. 
Since the CHR is co-aligned with the fundamental laser light, it is useful as an alignment aid in both the transverse and longitudinal directions.
Both optical beams are directed via a mirror transport system to a fast GaAs photodiode located outside of the accelerator shielding.

\begin{figure}[htb]
	\includegraphics[width=3.36in]{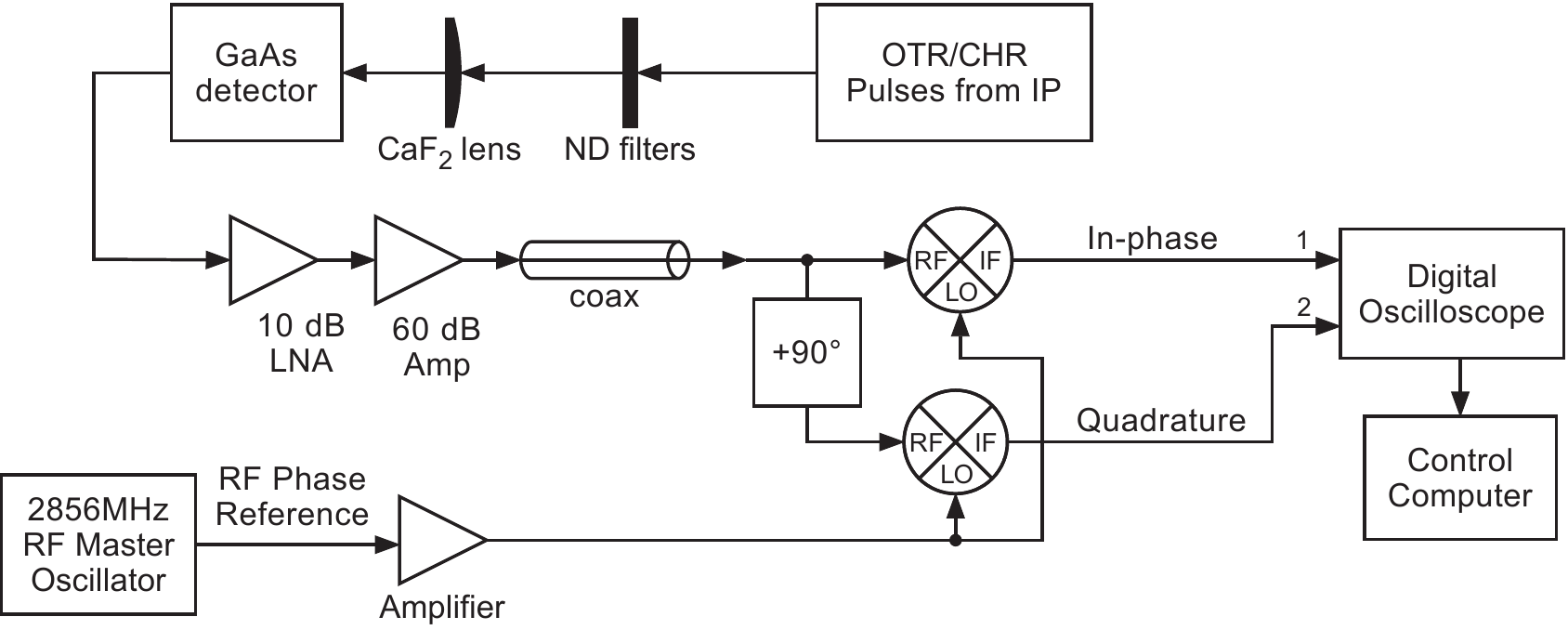}
	\caption{OTR and CHR signals detected by a fast GaAs photodiode are amplified and then demodulated and digitized by the complex receiver portion of the feed-forward amplitude and phase compensation system.  The quadrature and in-phase signal waveforms are used to reconstruct the amplitude and phase of the \RFG signals.
	\label{fig:feedforward}}
\end{figure}

A precision RF phase comparison apparatus has been designed and operated at the University of Hawai`i FEL Laboratory for the purpose of feed-forward stabilization of the linac's RF drive system\cite{hadmack2013ff}.  
While this system was intended for the analysis of amplitude and phase ripple on RF waveforms used for compensation, it has proved extraordinarily useful as a general purpose diagnostic for the comparison of the various S-band signals in the laboratory with the accelerator's master oscillator.  
Accepting the master oscillator as an absolute phase reference we can make independent but precisely calibrated measurements of the relative phases of the OTR and CHR light.  
A simplified block diagram of this phase measurement system is shown in Fig.~\ref{fig:feedforward}.

The phase and amplitude of an RF waveform are reconstructed using direct quadrature demodulation.
The input signal is $V_\textsc{s}^\textsc{i}(t) = A(t)\cos(\omega t + \phi(t))$, where $A(t)$ and $\phi(t)$ are the amplitude and phase modulations and $\omega = 2\pi(\RFG)$ is the RF carrier frequency.
The signal is split, with one branch delayed by \ang{90}, giving $V_\textsc{s}^\textsc{q}(t) = A(t)\sin(\omega t + \phi(t))$.
$V_\textsc{s}^\textsc{i}(t)$ and $V_\textsc{s}^\textsc{q}$ are individually mixed (multiplied) with the reference signal $V_R = \cos(\omega t)$ in double balanced mixers and then low-pass filtered to remove the second harmonics yielding:

\begin{align}
	I(t) &= \frac{A}{2} \cos\phi(t) \\
	Q(t) &= \frac{A}{2} \sin\phi(t)
\end{align}

The amplitude and phase can be reconstructed using:
\begin{eqnarray}
	A(t) = 2 \sqrt{I^2 + Q^2} \label{eqn:defamp}\\
	\phi(t) = \arctan (\frac{Q}{I}) \label{eqn:defphase}
\end{eqnarray}

Sampling the $I$ and $Q$ waveforms with an oscilloscope allows the amplitude and phase reconstruction to be performed by software in real time.


\section{Experiment}
\begin{figure}
	\includegraphics[width=3.36in]{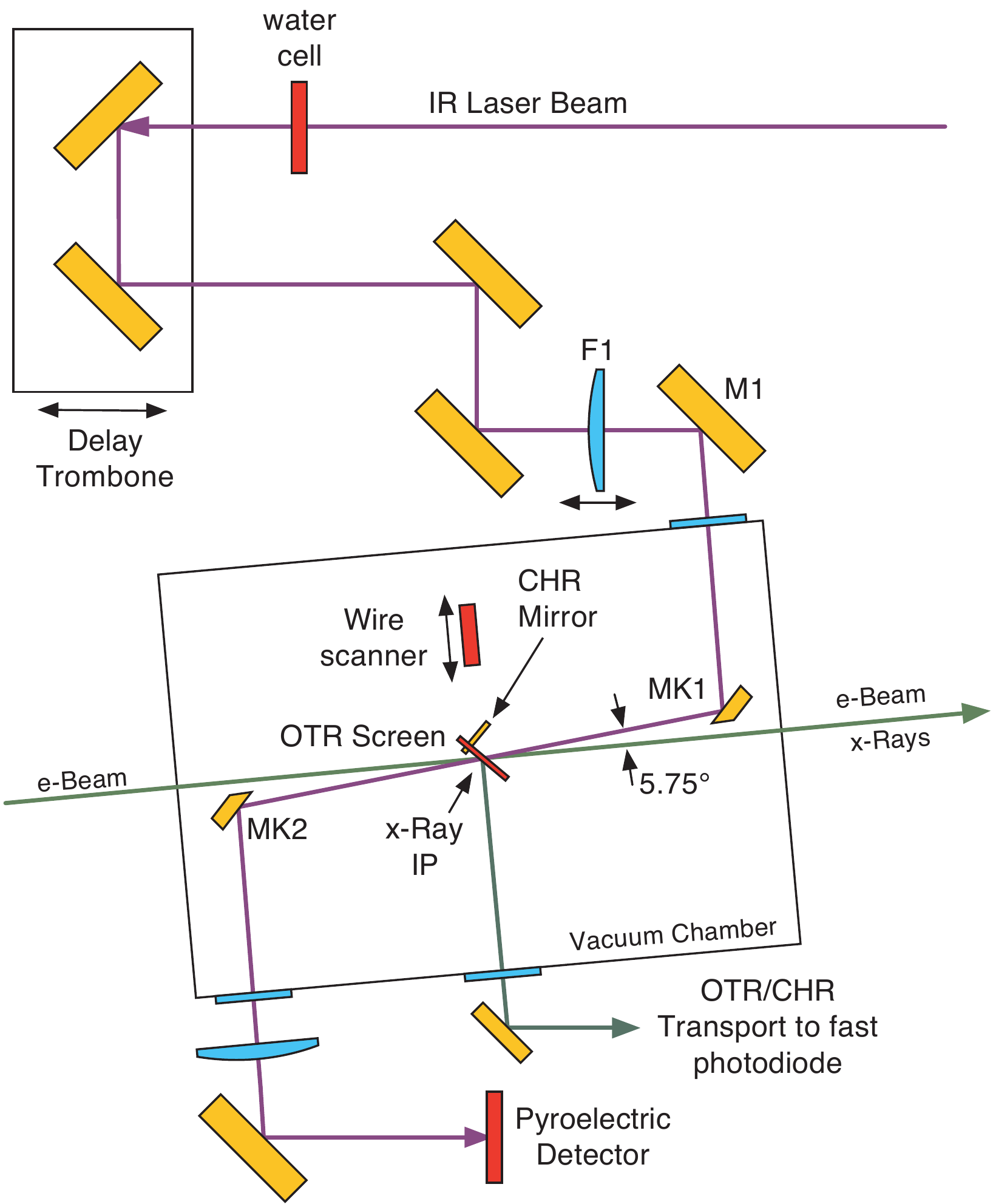}
	\caption{The inverse-Compton scattering apparatus consists of optics to delay, focus and steer the infrared laser beam to the interaction point located at the center of a vacuum chamber installed on the electron beamline.  Retractable kicker mirrors (MK1,2) enable injection of the laser for a nominal \ang{5.75} crossing with the electron beam.\label{fig:xrayoptics}}
\end{figure}

The x-ray interaction point is located in a vacuum chamber custom-built to house the optics and diagnostics required to facilitate efficient scattering.\cite{hadmack2012phd} 
Laser light from the FEL arrives at the interaction point via a vacuum transport line followed by a propagation of \SI{2}{\m} in air where the beam is aligned, focused, and injected into the interaction chamber for scattering and diagnostics.  
The complete optical system shown in Fig.~\ref{fig:xrayoptics} includes a ``trombone'' (optical delay line) used to synchronize the laser micro-pulse phase with the electron bunches.  
In normal operation the laser is aligned transversely and longitudinally to the IP with the mirror M1 and focused to a \SI{60}{\um} diameter waist by the \SI{250}{\mm} focal length lens F1.  
The spatial overlap is optimized and verified with the use of a scanning wire beam profiler.
A \SI{34}{\um} diameter carbon fibre is translated through both beams with the e-beam profile measured via the secondary emission current, and the laser profile measured by the wire occluded intensity.\cite{hadmack2013sca}

\begin{figure}
	\includegraphics[width=3.36in]{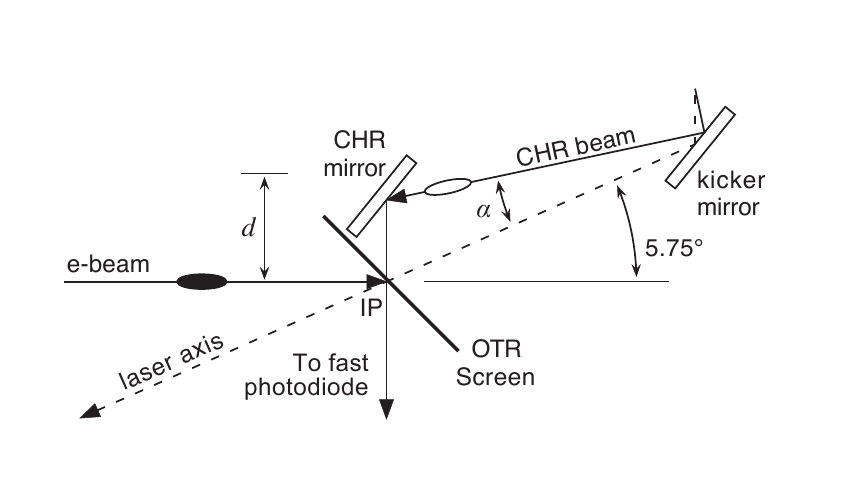}
	\caption{A small mirror can be installed offset from the electron beam in place of the OTR screen.  This mirror allows the CHR light from the FEL to be co-aligned with the OTR light from the e-beam without obstructing the e-beam.  The CHR pulses travel an additional distance $d$ relative to the laser beam when steered directly at the IP. \label{fig:cskicker}}
\end{figure}

\begin{figure}
	\includegraphics[width=2.7in]{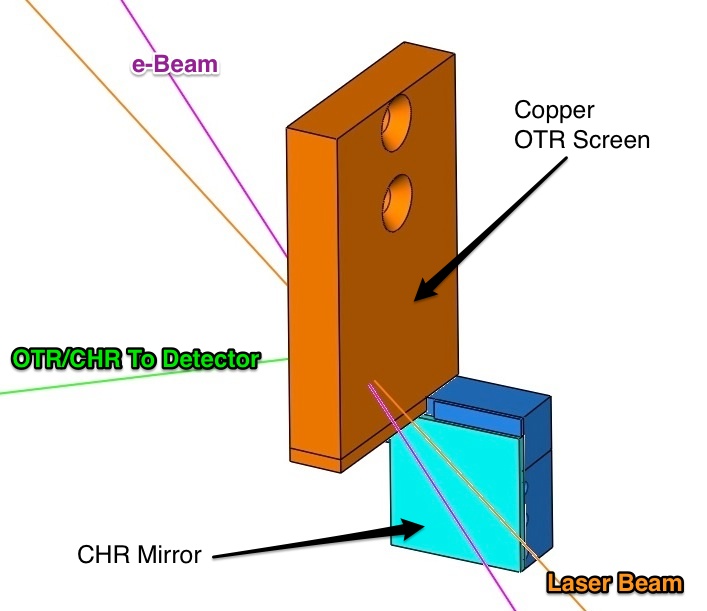}
	\caption{Insertable copper OTR screen inserted at the IP with the CHR mirror attached in the secondary insertion position.\label{fig:screen_photo}}
\end{figure}

A copper mirror oriented at \ang{45} can be inserted for imaging of the electron beam using OTR. 
However, since the laser beam is counter-propagating the same mirror cannot be used for extraction of the laser beam.  
Moreover, the unperturbed electron beam is required downstream of the IP for operation of the FEL.  
For the purpose of synchronization, a small silvered silicon mirror is mounted below the copper OTR mirror and inserted using the same pneumatic actuator but with a different insertion set point.  
This geometry is shown in Fig.~\ref{fig:cskicker}, and the dual mode OTR/CHR mirror assembly is shown in Fig.~\ref{fig:screen_photo}.
The mirror position is offset from the electron beam to provide a clearance of \SI{5}{\mm} when inserted (allowing for unimpeded FEL operation).
It is aligned such that deflection of the laser beam onto the mirror by $\alpha=\ang{1.75}$ using mirror M1 results in a CHR reflection co-aligned to the axis of the OTR light.  
A correction is applied to the measured laser phase to account for the path length difference of $d=\SI{6.78}{\mm}$ associated with this alignment.

It may be possible to eliminate the path length difference by using a thin silvered film of boron nitride, mounted directly below the OTR screen, capable of transmitting the e-beam to the laser with minimal perturbation.

Light from both the OTR and CHR is detected using a Hamamatsu G4176-03 GaAs photodiode with a \SI{30}{\ps} rise time.  
The G4176-03 is conveniently packaged with an SMA connector and can be directly attached to a bias-tee and an amplification chain composed of SMA connectorized microwave stock components from MiniCircuits.  
A gain of \SI{60}{\dB} is applied before sampling by the phase detection system.

Initial phase measurement attempts using a fast Varian crossed-field photomultiplier tube (PMT) showed that the phase of the RF signal depended strongly on light intensity, probably due to space charge effects in the tube, thus making precision phase measurements impractical with the PMT.  
Background radiation also confounded initial measurement attempts within the accelerator shielding, necessitating the installation of a \SI{7}{\m} long optical transport line allowing the detector to be used outside of the radiation shielding.  
The added system complexity was greatly offset by the ability to do hands-on alignment of the detector.

The spectral response of the G4176-03 photodiode is limited to visible light from \SIrange{400}{900}{\nm} and is well suited to the OTR light.  
While the laser light at \SI{3000}{\nm} is not detectable with GaAs, the beam contains many orders of co-propagating coherent harmonic undulator radiation\cite{bamford1990aa}; four of the harmonics overlap the sensitive range of the GaAs diode.  
To avoid damage to the diode and CHR mirror from the intense infrared laser beam, a water absorption cell is installed along the laser beamline for attenuation of the IR components while passing visible light.
Neutral density filters are also used to normalize the CHR intensity to that of the OTR. A correction for the group delay of the CHR due to these refractive elements is included in the synchronization calculation.  
Compared with a PMT, the semiconductor photodiode has the advantage that the detected signal phase is little affected by the intensity or wavelength of incident light.
While it would also be possible to compare the laser fundamental directly to the IR OTR light, the high cost of fast cryogenic mid-IR detectors makes this approach impractical compared to the method described here using co-propagating visible coherent harmonic radiation.


\section{Results}
For the synchronization measurements the free-electron laser was configured to provide a \SI{1.7}{\us} macro-pulse with an energy of \SI{3.6}{\milli\joule} at \SI{4500}{\nm}.  A \SI{37.1}{\MeV} electron beam with \SI{135}{\mA} over a \SI{2.5}{\us} macro-pulse was used.  Bright visible coherent harmonics of the laser light were detectable using the GaAs photodiode.

\begin{figure}
	\includegraphics[width=3.36in]{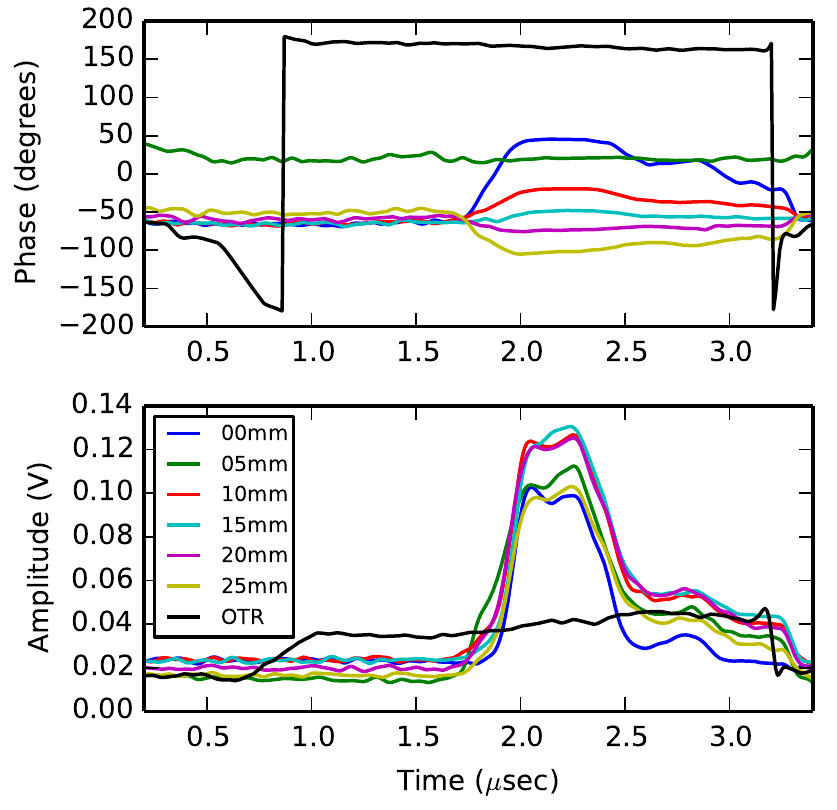}
	\caption{(color) Phase and amplitude of the \RFM component of the OTR light compared to the CHR light at various trombone positions.  Ten subsequent macro-pulses are averaged for each trace without background subtraction.
	\label{fig:rawwaveforms}}
\end{figure}

\begin{figure}
	\includegraphics[width=3.36in]{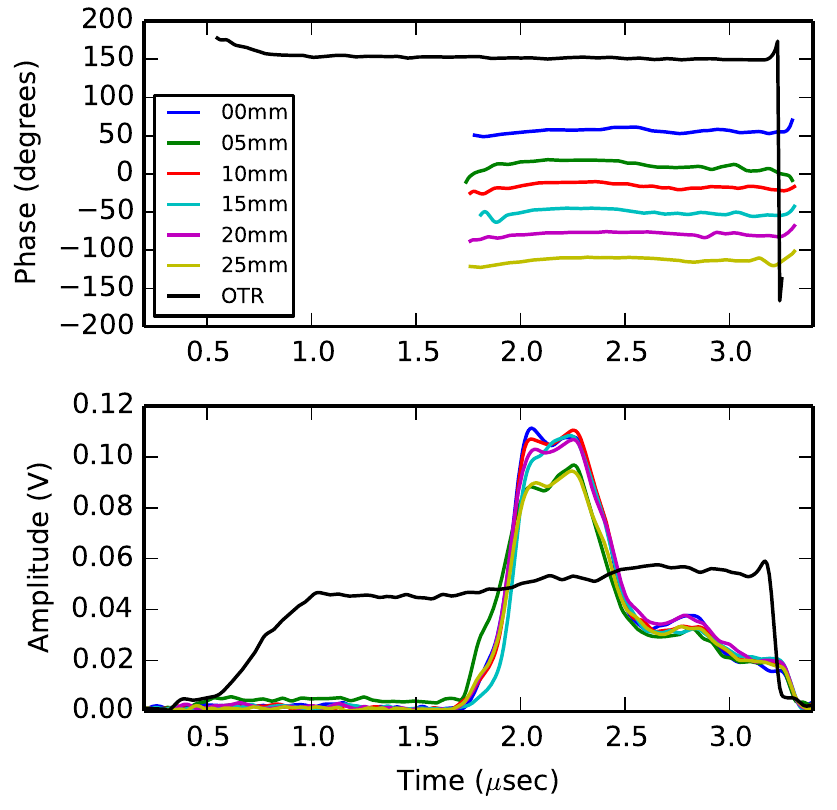}
	\caption{(color) Phase and amplitude of the \RFM component of the OTR light compared to the CHR light at various trombone positions.  The RF background pedestal has been subtracted from these data and the phase for signals with an amplitude (relative to background) less than \SI{6}{\mV} is excluded.  Each trace shown is the average of ten subsequent pulses with the background subtracted from each.\label{fig:apwaveforms}}
\end{figure}

The RF amplitude and phase compensation system\cite{hadmack2013ff} was used to record the complex waveforms of the \RFG component of the photodiode signal for the OTR light as well as the CHR light for six different trombone positions.  
Figure~\ref{fig:rawwaveforms} shows the raw demodulated waveforms without any background subtraction.  
The \SI{20}{\mV} pedestal seen in the lower amplitude plot is the result of a CW noise signal at \RFG picked up at some point in the signal chain.  
This noise signal varies in both amplitude and phase across successive macro-pulses but is essentially flat over the \SI{2.5}{\us} macro-pulse.  
The background signal phasor is estimated by averaging over the first \SI{500}{\ns} of each waveform and subtracting this complex value for the remainder of the waveform.  
Figure~\ref{fig:apwaveforms} shows each of the signals averaged over 10 macro-pulses with the background phasor subtracted for each pulse individually before averaging.  
Notice that the CHR amplitudes are now reasonably consistent while the phases are flat and evenly separated.

\begin{figure}
	\includegraphics[width=3.36in]{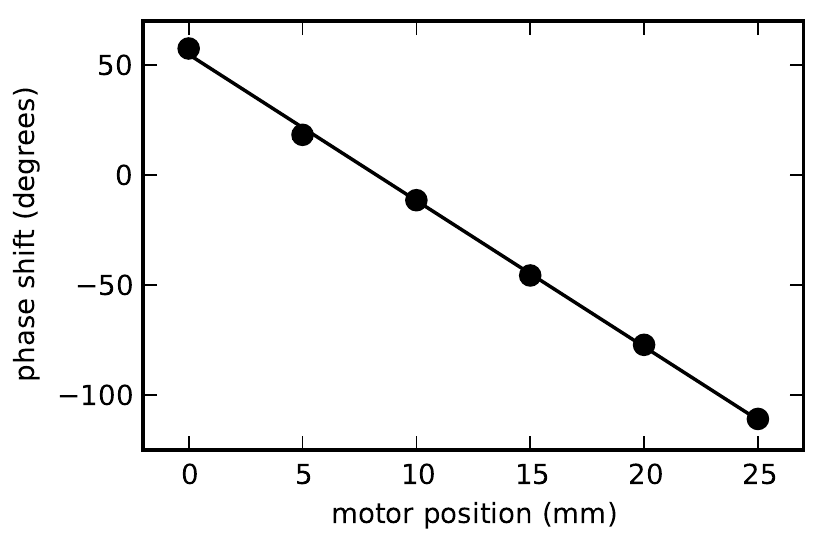}
	\caption{Linear dependance of the CHR phase on the trombone position providing the calibration necessary to synchronize the OTR and CHR pulses.\label{fig:trombonecal}}
\end{figure}

The phase $\phi$ of the CHR signal is averaged over the \SI{200}{\ns} region of peak signal amplitude and plotted in Fig.~\ref{fig:trombonecal} for each trombone position $z$.  
A linear regression $\phi_\textsc{chr} = \phi_0 + sz$ for these data provides the desired trombone calibration, yielding $s=\SI{-6.6 \pm 0.1}{\degree\per\mm}$ with $\phi_0 = \ang{54.9 \pm 1.6}$.
The OTR phase measured in the same interval is \ang{148.9 \pm 1.6}, and
\begin{equation}
	z_\textsc{tr} = \frac{\phi_\textsc{tr} - \phi_0}{s}
\end{equation}
gives a motor motion of \SI{-14.0 \pm 0.4}{\mm}.
To obtain a positive motion, \ang{360} is subtracted from $\phi_\text{TR}$ giving a \SI{+40.1 \pm 0.4}{\mm} trombone correction to synchronize the pulses at the detector.

The presence of four visible harmonics ranging from \SIrange{429}{750}{\nm} can contribute a phase error due to dispersive broadening of the pulse in the $\text{CaF}_2$ lenses used.  Assuming harmonics of equal intensity gives an upper limit phase error of \ang{0.88} corresponding to a trombone position error of less than \SI{0.13}{\mm}.  This dispersive error is not significant compared with the phase calibration error discussed above, but can be eliminated with the addition of a bandpass filter to select a single CHR harmonic.

The actual correction applied to the trombone motor contains an additional \SI{4.99}{\mm} offset, which accounts for the physical path length difference due to the CHR mirror offset shown in Fig.~\ref{fig:cskicker} as well as an offset to compensate for refraction in the filters and water cell which are not present during high power laser operation.
  
Note that if the trombone calibration slope $s$ is predicted based on the speed of light, then a value of \SI{6.86}{\degree\per\mm} is obtained, a 3.2\% deviation from the measured slope despite a stage position calibration error less than 1\%.
This error is most likely due to a deviation from the nominal \ang{90} phase difference between the in-phase and quadrature components prior to mixing with the RF reference. 
The demodulator in the present apparatus is constructed using matched lengths of miniature hardline coax. 
The observed 3.2\% scaling error, which corresponds to a \ang{3} error in the \ang{90} quadrature offset, would result from a \SI{0.8}{\mm} relative error in the cable lengths and is entirely possible. 
Such errors can be avoided in future designs with the installation of a mechanical line stretcher for calibration purposes. 
While this error precludes measurements of the absolute phase, it does not affect the relative phase measurements required in the present application, in which the measured slope can be precisely projected to obtain the correct synchronization phase.


\section{Conclusion}
An RF technique to synchronize the arrival of \si{\GHz} repetition rate trains of \si{\pico\second} electron bunches and laser pulses at the interaction point of an inverse-Compton scattering source of x-rays or gamma-rays was demonstrated.
This procedure allows the optical trombone to be adjusted to an uncertainty of less than \ang{2}, thus reducing the delay scan range for empirical optimization from \SI{52}{\mm} to \SI{0.3}{\mm}.  
Since a picosecond laser pulse occupies about \ang{1} in RF phase at S-band, use of this technique ensures that synchronization of the pulses will require a scan of less than twice the pulse length, greatly reducing parameter space to be optimized during the commissioning of the inverse-Compton source.
Furthermore, due to the shallow beam crossing angle and transverse beam size, the longitudinal distance over which collisions will occur is increased to several \si{mm}, thus increasing the ease of synchronization in the present application.


%
%

%

\begin{acknowledgments}
This work was funded under the Department of Homeland Security grant number 20120-DN-077-AR1045-02.
\end{acknowledgments}


%

\end{document}